\documentclass[preprint,flushrt]{aastex}
\usepackage{graphicx}
\newcommand{\eqb}{\begin{eqnarray}}
\newcommand{\eqe}{\end{eqnarray}}
\newcommand{\diff}{{\rm d}}
\newcommand{\mud}{\mu_+}
\newcommand{\pd}{p_+}
\newcommand{\urel}{u_{\rm rel}}
\newcommand{\grel}{\Gamma_{\rm rel}}
\newcommand{\adindex}{\hat{\gamma}}

\newcommand{\bperp}{B_{\mbox{\scriptsize$\perp$}}}
\newcommand{\lpar}{\ell_{\mbox{\scriptsize$\parallel$}}}
\newcommand{\lperp}{\ell_{\mbox{\scriptsize$\perp$}}}
\slugcomment{submitted to ApJ}
\shorttitle{Acceleration at ultrarelativistic shocks}
\shortauthors{Kirk et al.}

\begin{document}
%\thesaurus{}
\title{Particle acceleration at ultrarelativistic shocks:\\ 
an eigenfunction method}
\author{J. G. Kirk A. W. Guthmann}
\affil{Max-Planck-Institut f\"ur Kernphysik,
Postfach 10 39 80, 69029 Heidelberg, Germany}
\email{John.Kirk@mpi-hd.mpg.de Axel.Guthmann@mpi-hd.mpg.de}
\and
\author{Y. A. Gallant A. Achterberg} 
\affil{Sterrenkundig Instituut Utrecht, Postbus 80000, 3508 TA Utrecht, 
Netherlands}
\email{Y.A.Gallant@phys.uu.nl A.Achterberg@phys.uu.nl}
\begin{abstract}
We extend the eigenfunction method of computing the power-law spectrum of 
particles accelerated at a relativistic shock front 
 to apply to shocks of 
arbitrarily high Lorentz factor.
In agreement with the 
findings of Monte-Carlo simulations, we find the 
index of the power-law distribution of accelerated particles 
which undergo isotropic diffusion in angle at an 
ultrarelativistic, unmagnetized shock is 
$s=4.23\pm0.01$ (where $s=-\diff \ln f/\diff \ln p$ with $f$ the 
Lorentz invariant phase-space density and $p$ the 
momentum). This corresponds to a synchrotron index for uncooled electrons 
of $\alpha=0.62$ (taking cooling into account $\alpha=1.12$), 
where $\alpha=-\diff \ln F_\nu/\diff\ln \nu$, $F_\nu$ is the radiation flux and
$\nu$ the frequency. 
We also present an approximate analytic expression for 
the angular distribution of accelerated particles, which displays the 
effect of particle trapping by the shock: 
compared with the non-relativistic case the angular 
distribution is weighted more
towards the plane of the shock and away from its normal.
We investigate the sensitivity of our results to 
the transport properties of the particles 
and the presence of a magnetic field. 
Shocks in which the parameter $\sigma$ (the ratio of 
Poynting to kinetic energy flux) upstream 
is not small are less compressive and lead to larger values of $s$.  
\end{abstract}

\section{Introduction}

The theory of diffusive acceleration at shock fronts was first developed in
1977 -- for reviews see \citet{drury83}, 
\citet{jonesellison91} and
\citet{blandfordeichler87} --
and was quickly applied to many astrophysical problems. 
In its simplest form,
this theory assumes that accelerated particles diffuse in space 
upstream and downstream
of a discontinuity in the flow velocity of the plasma. This assumption 
requires the ratio of the plasma speed to the particle speed to be a small
quantity. The theory is, therefore, restricted to nonrelativistic flows.  
Although 
it was already well-known that relativistic flows exist and contain 
accelerated particles, it took some time for the theory to be 
extended into this domain. Until recently, the situation for relativistic shocks was that 
a semi-analytic eigenfunction method had been developed and was capable of 
computing the expected power-law index of accelerated particles
for shock fronts moving at 
Lorentz factors $\Gamma$ of up to roughly~5, 
assuming various models describing 
the way in which the particles diffuse in pitch angle in the 
upstream and downstream plasmas \citep{kirkschneider87,heavensdrury88}.  
This method also provided the
full angle and space dependence of the highly anisotropic 
distribution function.
Several sets of Monte-Carlo simulations had also been performed 
for the special case of isotropic pitch-angle diffusion, and these confirmed 
the semi-analytic results. 
For a review see \citet{kirkduffy99}.

Motivated mainly by developments in the field of gamma-ray bursts, 
Monte-Carlo simulations of acceleration at 
highly relativistic ($\Gamma>5$) shocks have recently
been presented by \citet{bednarzostrowski98}
and \citet{gallantetal98,gallantetal00}, 
indicating that 
the index $s$ of the power-law spectrum tends for large $\Gamma$ 
to a value close to 4.2, as had been speculated by 
\citet{heavensdrury88}.
In order to provide an independent 
check on the simulation results and to extend them to cover more general 
diffusion coefficients, we present in 
this paper a new eigenfunction method,
suitable for arbitrary shock speeds. Compared with the original method, the 
expansion of the distribution function in the new method 
converges much more rapidly, so 
that for most purposes only a single eigenfunction is required. This enables
rapid computation of $s$ over a wide range of parameter space and 
for a variety of angular diffusion coefficients. For highly relativistic
shocks, the angular distribution
at the shock front and at all points upstream is given by a simple analytic
expression, in which the details of the scattering and the downstream equation
of state enter only through the value of $s$.

It is at first sight not obvious that the first-order Fermi process will
operate at an ultrarelativistic shock front. Unless the system is
fine-tuned, such a shock front will be \lq superluminal\rq\ 
in which case the Fermi process requires the transport of
particles across field lines 
\citep{begelmankirk90,achterbergball94,michalekostrowski98}. 
This aspect of the problem and the way in which
we parameterize the cross-field transport is discussed in Sect.~\ref{transport}.
The new method is then described in Sect.~\ref{method}.
In Sect.~\ref{results} we first present results obtained
for isotropic angular diffusion at strong shocks in 
an ideal gas in which the magnetic field is dynamically unimportant.
We then consider the
spectrum produced by a decrease in the shock compression due to 
a finite 
magnetic field strength. 
Finally, we investigate the 
modifications introduced by anisotropic diffusion
coefficients. In Sect.~\ref{discuss} we summarize our results 
and briefly discuss their application to shock fronts in
astrophysical systems.

\section{Particle transport at ultrarelativistic shocks}
\label{transport}

In the case of a relativistic parallel shock as considered by
\citet{kirkschneider87}, the particle distribution is assumed 
gyrotropic (i.e., independent of gyro-phase) and particles 
diffuse in pitch angle. The particles are injected
with a Lorentz factor greater than that of the shock, but still small 
compared to that of the accelerated particles we are interested in.
They are treated as test-particles which do not affect the plasma conditions. 
The stationary transport equation satisfied in both the upstream 
and downstream regions is then
\eqb
\Gamma(u + \mu ){\partial f(p,\mu,z)\over\partial z}&=&{\partial\over\partial
\mu}\left[D_{\mu\mu}{\partial f(p,\mu,z)\over\partial\mu}\right]
\label{stattranseq}
\eqe
where $u$ is the speed of the plasma in units of the speed of light,
measured in the shock frame (in which the
distribution is assumed stationary),  
$\Gamma=(1-u^2)^{-1/2}$, $D_{\mu\mu}$ is the pitch-angle diffusion
coefficient and  $f(p,\mu,z)$ is the 
(Lorentz invariant) phase-space density,
assumed to be a function of the momentum $p$, the 
cosine of the pitch angle $\mu$ and 
a single Cartesian coordinate $z$. Both the momentum and 
pitch-angle variables are measured in the local rest frame of the plasma, 
whereas position is measured relative to the shock front in the 
shock frame.

However, a parallel shock front is a special case which is unlikely to be 
realized for high $\Gamma$. Upon boosting from the rest frame of the 
upstream plasma to the shock frame, the 
component of the magnetic field 
in the plane of the shock is amplified by a factor $\sim\Gamma$, whereas that along the shock
normal is conserved. Fine-tuning of the field
orientation upstream of the shock is required if it is to remain approximately
parallel i.e., sub-luminal \citep{kirkheavens89}. This, in turn,
means that no first order Fermi-type acceleration at all
is expected unless there exists a scattering mechanism which enables particles
to cross field lines \citep{begelmankirk90}.

In this paper, we assume such a scattering mechanism exists, and that it can
be described by an equation identical to Eq.~(\ref{stattranseq}), in which,
however, the variable $\mu$ no longer describes the (cosine of the) 
pitch angle of a particle, but instead the cosine of the angle between
a particle's direction of motion and the normal to the shock front
(which we call the \lq direction angle\rq).
Scattering of this kind can be produced by fluctuations in the magnetic field
which are of much shorter wavelength than the gyroradius of a
particle. The effect of these fluctuations cancels out in the standard
treatment of particle transport in a plasma, where one averages over many
gyrations \citep{luhmann76}. But, in the case of a
relativistic shock front, an upstream particle will perform only a fraction of a
gyration in between encounters with the front \citep{gallantachterberg99}, 
so that the standard treatment
does not apply. 
Downstream, it is widely assumed that a large amplitude turbulent magnetic
field 
is generated by the shock front. 
If we assume that this field reverses
direction on a length scale small compared to the gyro radius, then 
a section of the trajectory short enough to be considered unperturbed will
not resemble a helix, but rather a straight line. In each case it seems reasonable
to abandon the pitch-angle description altogether and consider instead diffusion in
direction angle.

Equation~(\ref{stattranseq}) with isotropic diffusion in
direction angle, i.e., $D_{\mu\mu}\propto (1-\mu^2)$ corresponds exactly to the model 
adopted in the Monte-Carlo simulations
of \citet{gallantetal00}
and approximately to the model used by \citet{bednarzostrowski98}. 
\citet{gallantachterberg99} use this 
model in the downstream region and,
in addition 
allow for deflection by a uniform magnetic field upstream of the shock. 

Isotropic diffusion in direction angle is, however, 
an idealization, since a preferred direction of the magnetic field -- if one
exists -- is likely
to influence the
transport.
It is widely thought that relativistic shocks 
will generate a highly tangled field. If this is seeded by the compression
of preexisting fluctuations in the upstream medium, then it might be
expected that the correlation length of the field in the $z$ direction normal 
to the shock front is much shorter than that in the $x$ and $y$ directions.
Consider the ideal case of 
a magnetic field $\vec{B}=(B_x(z),B_y(z), B_z)$ 
which is static in the plasma rest
frame and which fluctuates as a function of $z$ only. The statistical
properties of interest are specified as follows
\eqb
\left< B_x(z')B_x(z+z')\right>
&=&
\left< B_y(z')B_y(z+z')\right>
\,\equiv\,  S(|z|)\\
\noalign{with}
S(z)&=&0\ {\rm for }\ z>\lpar\\
\noalign{The average of the square of the magnetic field 
$\left<{\bperp}^2\right>$ is defined by}
\int_{-\infty}^{+\infty}\diff z S(|z|)&\equiv&\left< {\bperp}^2\right>\lpar
\label{statistical}
\eqe
where $\left<\dots\right>$ denotes an ensemble average.

Integrating the equation of motion of a particle of charge $e$, mass $m$,
velocity $v$ and
Lorentz factor $\gamma$ moving in the $x$-$z$ plane with 
direction angle $\arccos\mu$, we find the
change in $\mu$ after an elapsed time $t$ to be:
\eqb
\Delta\mu&=&{e\over\gamma mc}
\int_0^t\diff t' \sqrt{1-\mu^2}B_y(z(t'))
\eqe
In the usual way -- see, for example \citet{ichimaru73} -- we compute the
diffusion coefficient by integrating over a time $\tau$ long compared to the
correlation time of the field (in this case $\lpar/\mu v$) but short
enough to allow us to use the unperturbed trajectory: $\mu=\,$constant, 
$x=\sqrt{1-\mu^2}vt$, $y=0$, $z=\mu v t$. Taking an ensemble average, we
find 
\eqb
{\left<\Delta \mu^2\right>\over \tau }
&=&{e^2\over\gamma^2 m^2c^2}{1-\mu^2\over|\mu| v}\left<{\bperp}^2\right>\lpar
\eqe
Clearly, this expression is unphysical in the neighborhood of $\mu=0$, because
we have assumed an infinite correlation length in the $x$ and $y$ directions.
Eliminating this unphysical behavior we write for
the diffusion coefficient
\eqb
D_{\mu\mu}&=&{e^2\over\gamma^2 v m^2 c^2}{1-\mu^2\over\sqrt{\mu^2+
\left(\lpar/\lperp\right)^2}}
\left<{\bperp}^2\right>\lpar
\label{tangled}
\eqe
where $\lpar$ and $\lperp$ are related to the correlation
lengths parallel and  perpendicular to the shock normal, respectively.

The diffusion coefficient in Eq.~(\ref{tangled}) is plausible in the 
downstream medium, which
is compressed along the shock normal. In the upstream medium this should not
be the case. However, especially for highly relativistic shocks, it is 
immaterial which diffusion coefficient is used there, since only that part
of the function $D_{\mu\mu}$ is of importance which lies close to $\mu=-1$.
In Sect.~\ref{anisotropicdiff}
we investigate the effect
of the strongly anisotropic diffusion coefficient given by
Eq.~(\ref{tangled})
on the acceleration process.

\section{The eigenfunction method}
\label{method}
Consider now an infinite plane shock front, 
so that in the upstream region $z<0$ and $u=u_-$ and in the downstream region 
$z>0$ and $u=u_+$, with both $u_\pm$ positive and $u_->u_+$.

Since there is no intrinsic momentum scale in Eq.~(\ref{stattranseq}), 
it follows that only the boundary conditions can introduce one into the solution.
If the space boundaries are far from the shock, and the lower momentum
boundary (the injection momentum) is well below the range of interest,
the intrinsic spectrum of particles accelerated by the shock is a power
law $f\propto p^{-s}$.
The problem is to find the index $s$, which depends on
the plasma speeds (upstream and downstream) and the function 
$D_{\mu\mu}$. 

Separating the variables $p$, $z$ and $\mu$ in Eq.~(\ref{stattranseq})
results in an eigenvalue problem for the angular 
part of the distribution:
\eqb
{\partial\over\partial \mu}
\left[ D_{\mu\mu}{\partial\over\partial \mu}
Q_i(\mu)\right]
&= &\Lambda_i(u+\mu) Q_i(\mu)
\label{eigvalprob}
\enspace,
\eqe
with the boundary conditions that the 
eigenfunctions be regular at the singular points $\mu=\pm1$.
The general solution for the distribution function is then written
\eqb
f&=&\sum_{i=-\infty}^{\infty}
a_ip^{-s}Q_i(\mu){\rm exp}(\Lambda_i z/\Gamma)
\label{eq4.1}
\eqe
and is valid in each half-space $z\ge0$ and $z\le0$.
At the shock front ($z=0$) Liouville's Theorem and the appropriate Lorentz transformations
for the particle momentum and direction relate the upstream and downstream distributions.
The eigenfunctions $Q_i$ in Eqs.~(\ref{eigvalprob}) and (\ref{eq4.1}) 
fall into two families: those
with positive eigenvalue, $\Lambda_i>0$, which are labeled $i=1,\dots\infty$ and those with
negative eigenvalue, labeled $i=-1,\dots-\infty$. In addition, there is a
special eigenfunction $Q=\,$constant, $\Lambda=0$ for which we choose $i=0$.
The eigenfunctions obey an orthogonality relation
\eqb
  \int_{-1}^1(u+\mu)Q_iQ_jd\mu=0  \qquad   \forall i\ne j \enspace.
 \label{orthorel}
\eqe

Imposing the condition that the distribution function far downstream (i.e.\ as $z\rightarrow\infty$) 
does not diverge, the distribution function for
$z>0$, simplifies to
\eqb
 f^+=\sum_{i=-\infty}^0a_i^{+}\left(p_+\right)^{-s}Q_i^{+}\exp(\Lambda_i^{+}z/\Gamma^+)\enspace.
\label{downdist}
\eqe
where the super- (or sub-)script \lq\lq$+$\rq\rq\ denotes a downstream quantity.
Far upstream ($z\rightarrow -\infty$) the distribution function should not only be
regular, but there should also be no incoming accelerated particles. Thus we can write
\eqb
 f^-=\sum_{j=1}^{\infty}a_j^-\left(p_-\right)^{-s}
Q_i^-\exp(\Lambda_j^{-}z/\Gamma^-)\enspace.
\eqe
where the super- (or sub-)script \lq\lq$-$\rq\rq\ denotes an upstream quantity.

In the original method \citep{kirkschneider87}, 
the downstream distribution function $f^+$ was expanded up to 
the term
$i=-N$ in Eq.~(\ref{downdist}). Then, the condition that the distribution vanish for 
$z\rightarrow-\infty$ was satisfied approximately by using the orthogonality 
relation (\ref{orthorel}) to project $f^-$ onto the 
eigenfunctions $Q_i^{-}$  and demanding that the terms $i=-N,\dots 0$ should vanish. 
This resulted in a set of $N+1$ linear homogeneous algebraic
equations for the $a_i^{+}$. The vanishing of the determinant of this system 
was used to find $s$.
Thus, only those eigenfunctions were required for which $i\le0$, 
which (except for $i=0$) are oscillatory 
in the range in which particles move 
in the same direction as the flow as seen 
in the shock frame i.e., $-u_\pm<\mu_\pm<1$
These were evaluated using a Galerkin method.

In the new method, we adopt the \lq mirror image\rq\ of this approach, expanding
the upstream distribution function to $N$ terms as an ansatz
\eqb
f^-&=&p_-^{-s}\sum^{i=N}_1
a_i^-
Q_i^{-}(\mu_-){\rm exp}(\Lambda^{-}_i z/\Gamma_-),
\label{new}
\enspace\eqe
for $z\le0$, which fulfills the upstream boundary condition 
at $z\rightarrow-\infty$. We then 
determine
$s$ by projecting onto the functions $Q_i^{+}$ for $i=1,\dots N$ and solving
the resulting $N$ homogeneous equations. In this case, only those
eigenfunctions are required for which $i\ge1$. Except for $i=1$
these are oscillatory in the interval $-1<\mu<-u$. At ultrarelativistic shocks,
$u_-\rightarrow1$, so that the oscillatory region for upstream eigenfunctions 
becomes small. Although the
eigenfunctions are then difficult to evaluate using Galerkin techniques, 
the direct numerical integration via shooting method and Pr\"ufer transformation 
used by \citet{heavensdrury88} works well.
Furthermore, an analytic expression for these eigenvalues and eigenfunctions 
has been given in the limit $u\rightarrow1$
\citep{kirkschneider89} and it turns out that 
the expansion in Eq.~(\ref{new}) converges much more rapidly than that in
Eq.~(\ref{downdist}).

\begin{figure}
\plotone{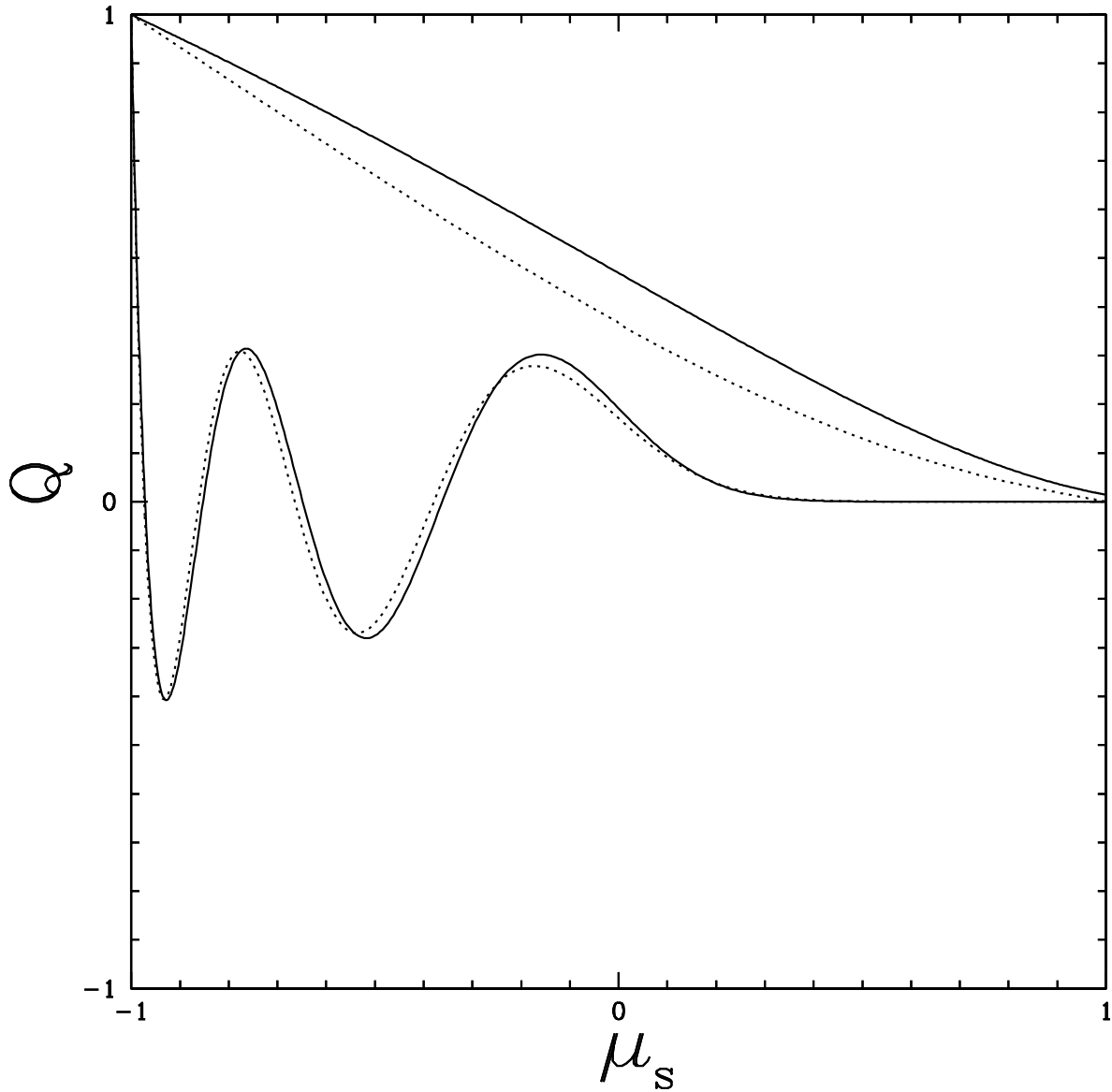}
\caption{\protect\label{eigenfunctions}
The eigenfunctions $i=1$ and $i=5$ 
for isotropic diffusion in direction angle 
as a function of direction angle measured in the
shock frame. 
The solid line depicts the results of a numerical 
integration using the Pr\"ufer transformation
for a plasma speed $u=0.5$, 
the dotted line shows the 
asymptotic expressions of Eq.~(\protect\ref{asymptoticeigf})
for $u\rightarrow1$
}
\end{figure}
An example of the eigenfunctions 
for isotropic diffusion in direction angle $D_{\mu\mu}=(1-\mu^2)D(\mu)$, 
with $D(\mu)=\,$constant, 
is shown in Fig.~\ref{eigenfunctions},
where the  abscissa is chosen to be 
the (cosine of the) direction angle $\mu_{\rm s}$
measured in the frame which projects $\mu=-u$ onto
$\mu_s=0$, thus stretching the oscillatory range. 
For $u=u_\pm$, this reference frame corresponds to the shock
frame for $Q^{\pm}$.
The $i$'th eigenfunction possesses $i-1$ roots 
in the interval $-1<\mu_{\rm s}<0$, corresponding to particles streaming 
against the flow. Only 
the first eigenfunction $i=1$ is
positive definite. According to \citet{kirkschneider89},
the eigenfunctions for a general diffusion coefficient with $D(\mu)$ 
are given in the limit $u\rightarrow1$ by the expressions
\eqb
Q^\infty_i(y)&=&\exp\left[-\epsilon y\sqrt{\Lambda_i^\infty /D(-1)}\right]
\sum_{n=0}^{i-1}c_ny^n\enspace,
\label{asymptoticeigf}
\eqe
where $\epsilon=1-u$ and $y=(1+\mu)/\epsilon$. 
The asymptotic expression for the 
eigenvalues is 
\eqb
\Lambda_i^\infty=(2i-1)^2 D(-1)/\epsilon^2
\label{asymptoticeigv}
\eqe
and the coefficients are determined by the recursion formula
$c_{n+1}=2(2i-1)(i-n-1)c_n/(n+1)^2$.
Figure~\ref{eigenfunctions} shows both the numerically calculated eigenfunctions 
(using the
Pr\"ufer transformation and shooting method) 
for $u=0.5$ and $D_{\mu\mu}\propto(1-\mu^2)$ and the asymptotic expressions
of Eq.~(\ref{asymptoticeigf}). For larger values of $u$, the numerical results
for both eigenfunctions and eigenvalues
are well approximated by the asymptotic expressions.
 
Matching the distributions across the shock front i.e., demanding
$f^+(\pd,\mud,0)=f^-(p_-,\mu_-,0)$
involves the transformations
\eqb
\pd&=&\grel p_- (1+\urel\mu_-)
\nonumber\\
\mud&=&(\mu_-+\urel)/(1+\urel\mu_-)
\eqe
where the relative velocity of the upstream medium with respect to the downstream one
is
\eqb
\urel\,=\,(u_--u_+)/(1-u_-u_+)
\nonumber
\eqe
and $\grel=(1-\urel)^{-1/2}$.

The projection of the expansion in Eq.~(\ref{new}) onto the downstream
eigenfunctions yields
\eqb
\sum_{j=1}^{j=N} S_{ij}a_j^-&=&0
\eqe
where the matrix $S_{ij}$ is given by
\eqb
S_{ij}&=&
\int_{-1}^{+1}\diff\mud (u_++\mud)(1+\urel\mu_-)^s Q^-_i(\mu_-)Q^+_j(\mud)
\label{matrixequation}
\eqe
The index $s$ is found from the condition $|S_{ij}|=0$ and the corresponding 
distribution function is given by Eq.~(\ref{new}) with those $a_i^-$ 
which lie in the null-space of $S$.

It is interesting to note that in the limit $u_-\rightarrow1$, the matrix
elements in Eq.~(\ref{matrixequation}) may be written
\eqb
S_{ij}&\rightarrow&
\int_0^\infty
\diff y Q^\infty_i(y)(y-1)(y+2)^{-s}Q^+_j(\mud)
\label{limitingmatrix}
\eqe
with $\mud=(y-2)/(y+2)$, and $Q^\infty(y)$ given by Eq.~(\ref{asymptoticeigf}).

\section{Results}
\label{results}
\subsection{Isotropic diffusion in angle}
\label{resultsisotropic}
First of all, we consider a plasma in which the particle transport is
described by isotropic diffusion in direction angle 
($D_{\mu\mu}=(1-\mu^2)D$ with $D$ constant)
and  the magnetic field is dynamically unimportant. 
In this case, the J\"uttner/Synge
equation of state describes the plasma, and the jump conditions across the
shock front must in general be evaluated numerically. However, in the case of
a strong shock, a useful analytic expression can be found using an 
approximate equation of
state in which the downstream ratio of specific heats
$\adindex$ is prescribed \citep{blandfordmckee77}:
\eqb
w_+/\rho_+&=&{\adindex}(\Gamma_{\rm rel}-1)+1
\nonumber\\
\Gamma_-^2&=&{ (w_+/\rho_+)^2(\Gamma_{\rm rel}+1)\over
{\adindex}(2-{\adindex})(\Gamma_{\rm rel}-1)+2}
\label{fixbland}
\eqe
where $w$ is the proper enthalpy density and $\rho$ is
the proper (rest-mass)
density.

Another straightforward case is that of 
a relativistic gas both up and downstream,
in which case one can derive
\eqb
 u_-u_+=1/3
 \label{relthird}
\eqe 
This situation is perhaps 
less likely to occur in practice, since it describes a shock front propagating into a 
medium in which the rest-mass energy density is negligible, even though the pressure may be important. 
In the limit $\Gamma_-\rightarrow\infty$ both (\ref{fixbland}) and (\ref{relthird})
give $u_+\rightarrow 1/3$, as must any physically acceptable 
equation of state describing an unmagnetized gas.

\begin{figure}
\plotone{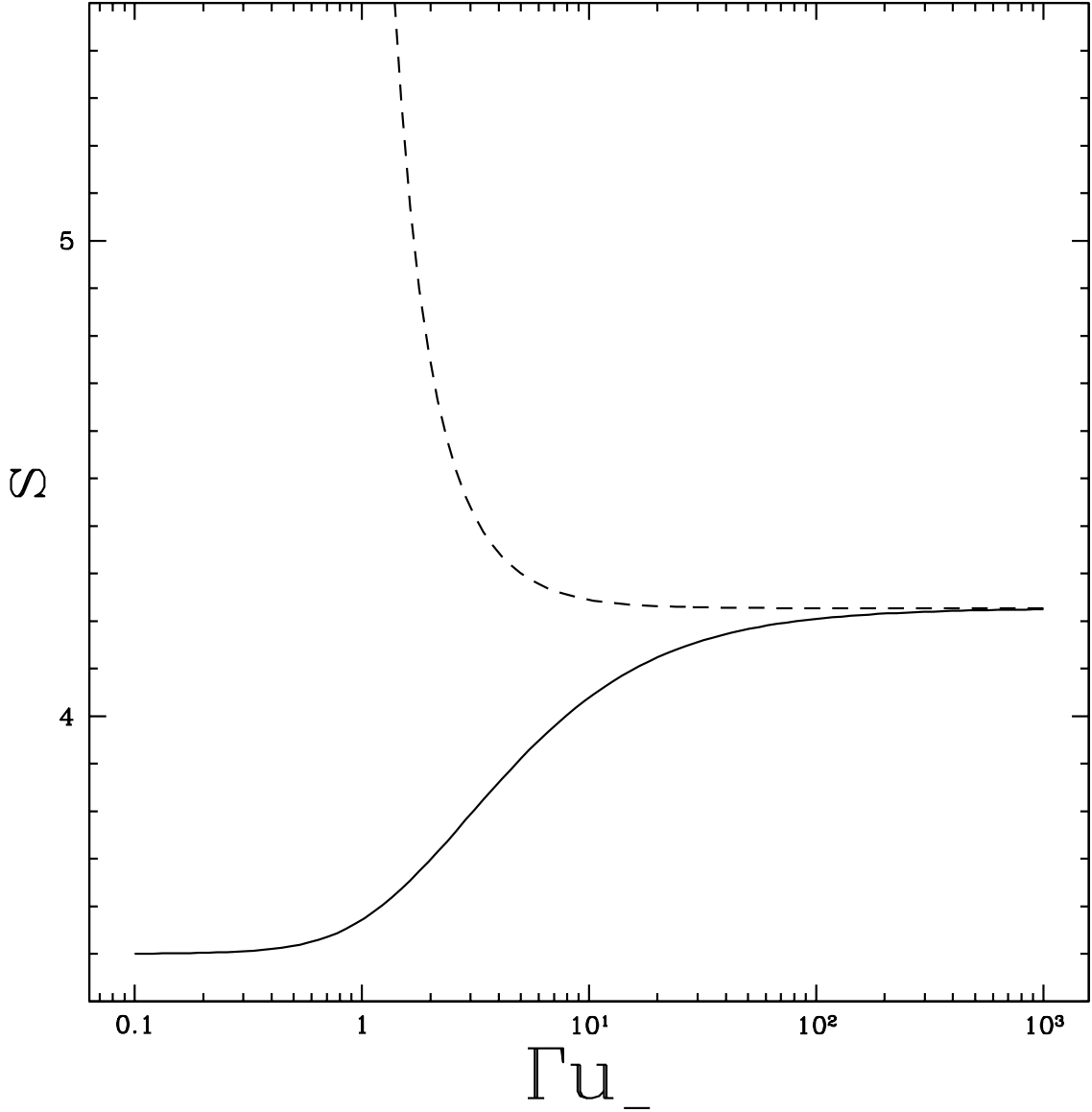}
\caption{\protect\label{fixedgamma}
The power-law index $s$ as a function of the 
spatial component of the upstream shock speed $\Gamma_-u_-$ for a strong shock
with fixed adiabatic index 
$\protect\adindex=4/3$ (solid line) and for a shock in a relativistic gas
(Eq.~\protect\ref{relthird}). 
The apparent asymptotic value of $s=4.23$ agrees with 
explicit computations performed in the limit $u_-\rightarrow1$.}
\end{figure}
  
Results obtained using these jump conditions are presented in 
Fig~\ref{fixedgamma}. The power-law index $s$ clearly tends to a limiting 
value as $u_-\rightarrow1$. This limit agrees with the value we find
from the asymptotic expression in Eq.~(\ref{limitingmatrix}):
\eqb
s&=&4.23\pm0.01
\eqe
where the errors quoted are our rough estimates of the accuracy of the 
numerical root finding algorithm, and the truncation error arising from 
the expansion of the distribution.
\begin{figure}
\plottwo{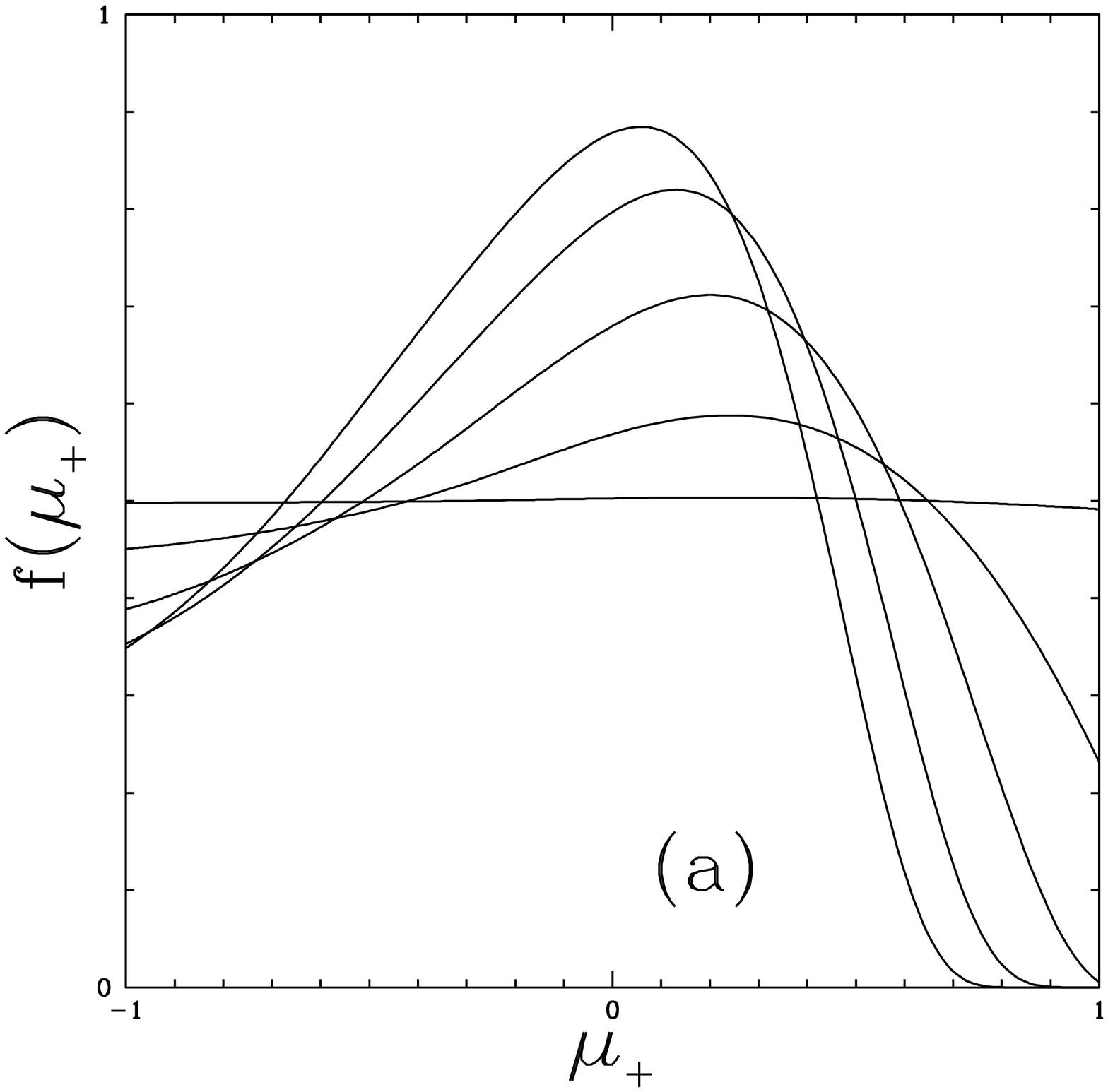}{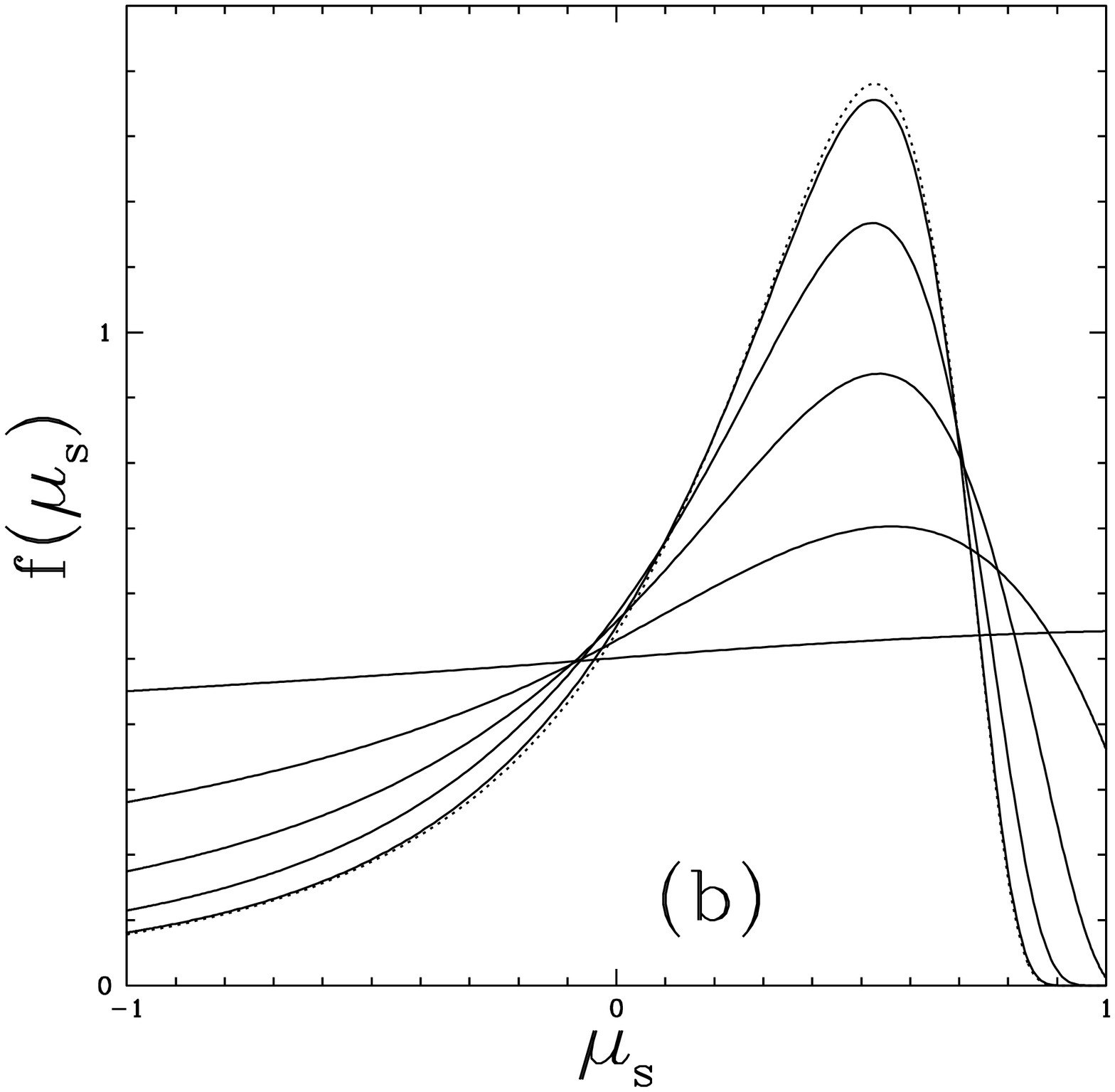}
\caption{\protect\label{angulardist}
The angular distribution of particles at 
the shock front (a) as seen in the downstream rest frame
as a function of the direction angle $\mu_+$ and
(b) as seen in the shock rest frame, as a function of 
the direction angle $\mu_{\rm s}$. Solid line curves are plotted for 
five different values of the upstream speed: $\Gamma_-u_-=0.1$, $0.5$, $1$, 
$2$, and $10$. The dotted line in (b) 
shows the asymptotic angular distribution 
given by Eq.~(\ref{asympdist}) for 
$\Gamma_-u_-\rightarrow\infty$ and $s=4.23$. 
The almost isotropic distribution at low $u_-$ 
gives the almost horizontal lines.
The peak in the flux of particles  
is more pronounced at higher $\Gamma_-u_-$. The jump conditions used 
are those of a strong shock with the 
J\"uttner/Synge equation of state downstream.}
\end{figure}

The corresponding angular distribution 
at the shock front is shown in Fig.~\ref{angulardist}
for several different values of $u_-$.  
In the nonrelativistic limit, the distribution
function is approximately isotropic and this angular distribution 
is an almost horizontal line. As the upstream speed increases, 
a pronounced hole appears in the distribution at the point where 
particles enter the downstream region along the shock normal.
Because upstream particles are caught by the shock before they can be substantially deflected, 
particles overtaken by a relativistic shock never re-enter the shock front along the normal.
To a lesser extent, those particles 
re-entering the shock from the downstream side
are also depleted 
at more relativistic shocks. These properties 
express the fact that the 
return probability of a particle leaving the shock by moving into 
the downstream region depends on the direction angle. The 
shock \lq captures\rq\ preferentially 
those particles of $\mu_{\rm s}\lesssim 0.6$. 
The absence of large deflections upstream 
ensures the energy gain per crossing remains modest after the first 
shock encounter, at which it may reach a factor 
$\Gamma_-^2$ \citep{vietri95,gallantachterberg99}.

We checked the convergence properties of our results by 
changing the number of terms in the expansion.
For $N>5$ we found convergence to within 
the numerical noise associated with the integration and root finding 
routines. An accuracy of better
than 10\% in all cases is already obtained by using just 
a single term in the expansion, i.e., $N=1$. 
Except in the case of anisotropic diffusion 
(see Sect.~\ref{anisotropicdiff}), the 
angular distribution was also found to be well represented by the first
eigenfunction when transformed 
to the shock frame.
For relativistic shocks ($u_->0.5$) this distribution can
be approximated by the expression given in Eq.~(\ref{asymptoticeigf}).
When written in the shock frame, the distribution then becomes:
\eqb
f_{\rm s}&\propto& \left(1-\mu_{\rm s}u_-\right)^{-s}
\exp\left(-{1+\mu_{\rm s}\over 1-u_-\mu_{\rm s}}\right)
\label{asympdist}
\eqe

\begin{figure}
\plotone{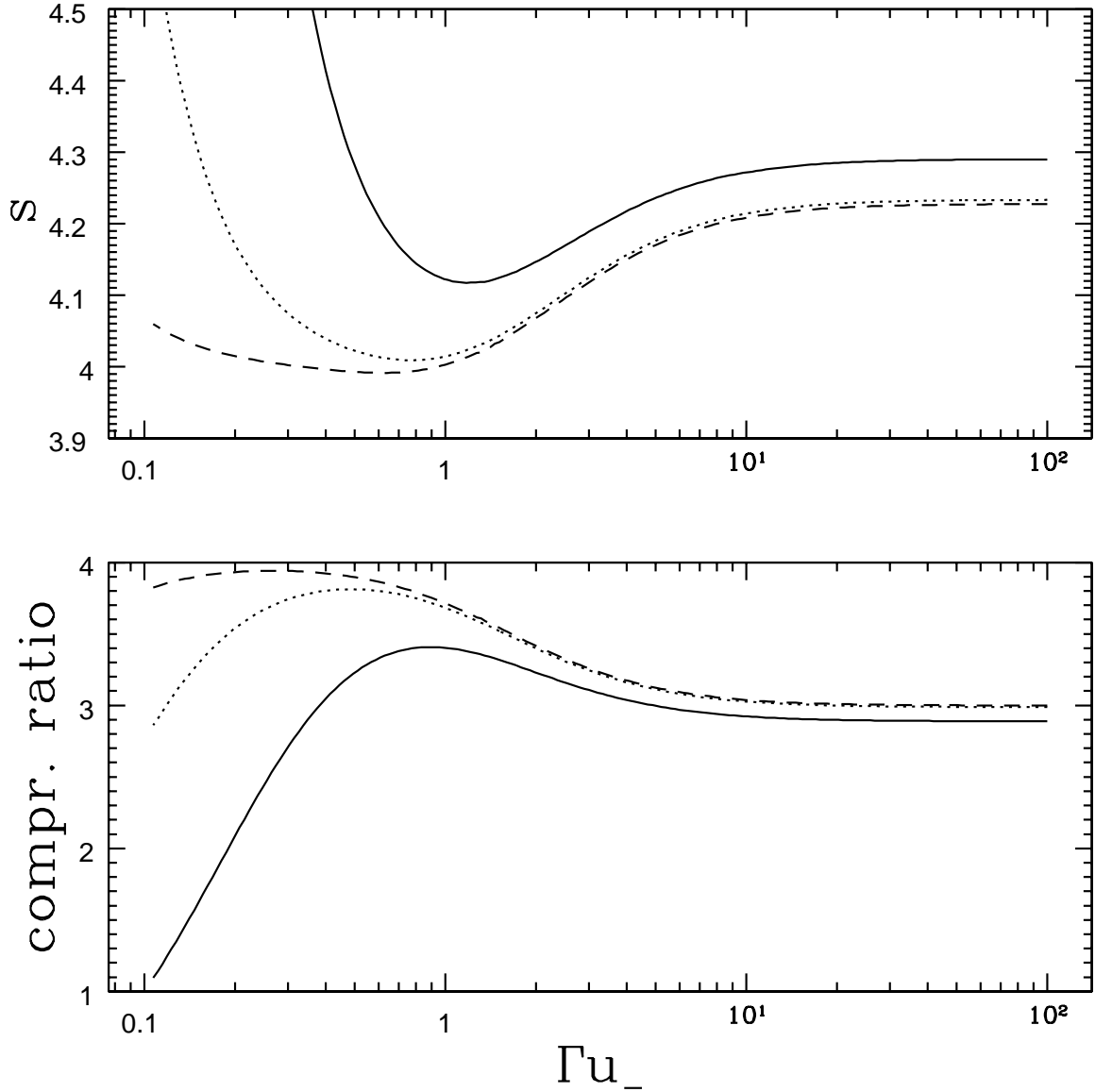}
\caption{\protect\label{relativistics}
The compression ratio and 
power-law index $s$ of accelerated particles as 
functions of the spatial component of the upstream four velocity at a 
strong shock in a magnetized plasma. The curves correspond to 
    $\sigma =10^{-2}$ (solid line) 
    $10^{-3}$ (dotted line) and 
    $10^{-4}$ (dashed line). The J\"uttner/Synge equation of state
was used and a strong shock assumed.  }
\end{figure}

\subsection{Strongly magnetized plasma}
\label{sectionmagnetised}
The strength of the magnetic 
field in the upstream or downstream plasma can be conveniently 
expressed in terms of the Lorentz invariant $\sigma$-parameter. 
This is defined as
\eqb
\sigma&=&{ B^2/4\pi w}
\eqe
\citep{kirkduffy99} where $B$ is the magnetic field
and $w$ the enthalpy density, both measured in the local rest frame. 
In the case of a cold MHD flow 
carrying a magnetic field perpendicular to the flow direction, 
this parameter describes the ratio of the Poynting flux to the 
energy flux density carried by the particles 
\citep{kennelcoroniti84a,michelli99}.
A shock moving at speed $u_-$ into the upstream 
medium with a magnetic field perpendicular to the shock normal then has a 
fast magnetosonic Mach number $M_{\rm fast}$
(the ratio of $u_-$ to the 
fast magnetosonic wave speed) given by 
\eqb
M_{\rm fast}&=&u_-\sqrt{1+\sigma\over\sigma+v^2_{\rm s}}
\label{alfvmachnumber}
\eqe
where $v_{\rm s}$ is the sound speed in the plasma, in units of the speed of
light.

At nonrelativistic shocks, it is usually assumed that Alfv\'en waves are
responsible for scattering and that they propagate away from
the shock front, since they are thought to be generated by the streaming of
the accelerated particles. This results in an effective reduction in the 
velocity discontinuity experienced by the particles by an amount which
depends on the 
magnetic field strength, and can have a significant 
influence on the predicted power-law index. In our case, however, where 
direction-angle scattering rather than pitch-angle scattering is important,
there is no clear connection between the average magnetic field strength and 
the speed of
the scattering centers. Nor is it obvious 
that the accelerated particles are responsible for generating the
fluctuations. In view of this, we make the simple assumption that they are 
frozen into the background plasma.

Nevertheless, the magnetic field can influence both the angular distribution and
the spectrum of accelerated particles by changing the jump 
conditions across the shock front. These can be computed 
numerically using a straightforward algorithm 
\citep{majoranaanile87,kirkduffy99}, into which one can
incorporate the full J\"uttner/Synge equation of 
state. Figure \ref{relativistics} shows results 
obtained using this equation 
of state for cold upstream plasma ($v_{\rm s}=0$)
for a range of $\sigma$ values. 

The general effect of finite $\sigma$ is to reduce the compression ratio. 
This leads to steeper spectra for the accelerated particles, as
found by \citet{ballardheavens91}. Note 
that, in accordance with Eq.~(\ref{alfvmachnumber}), the compression in 
the nonrelativistic limit does not tend to the unmagnetized value, but 
remains substantially weaker, finally disappearing when the shock speed 
reaches the fast magnetosonic speed ($M_{\rm fast}=1$) at
$u_-=\sqrt{\sigma/(1+\sigma)}$.

\begin{figure}
\plottwo{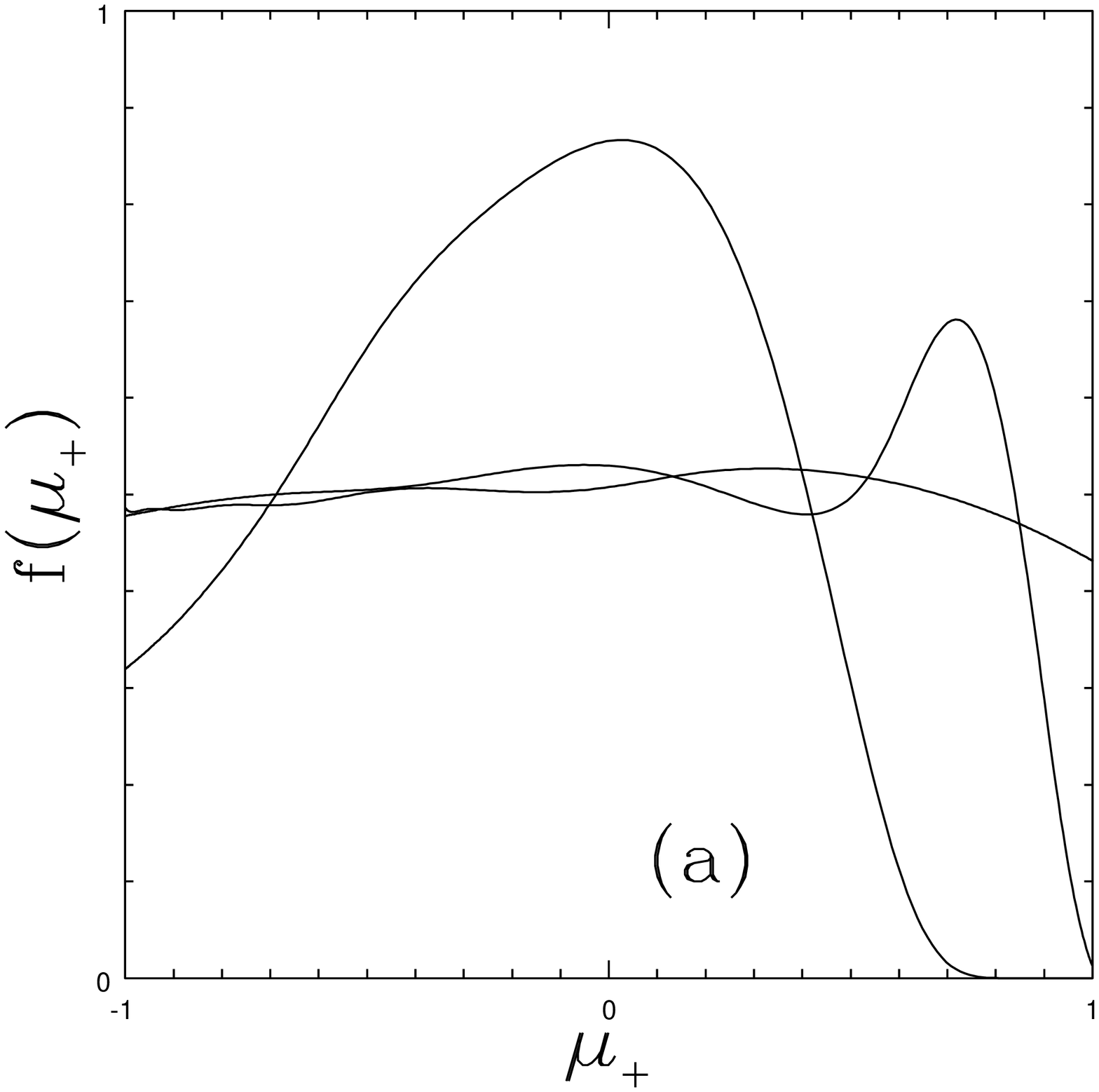}{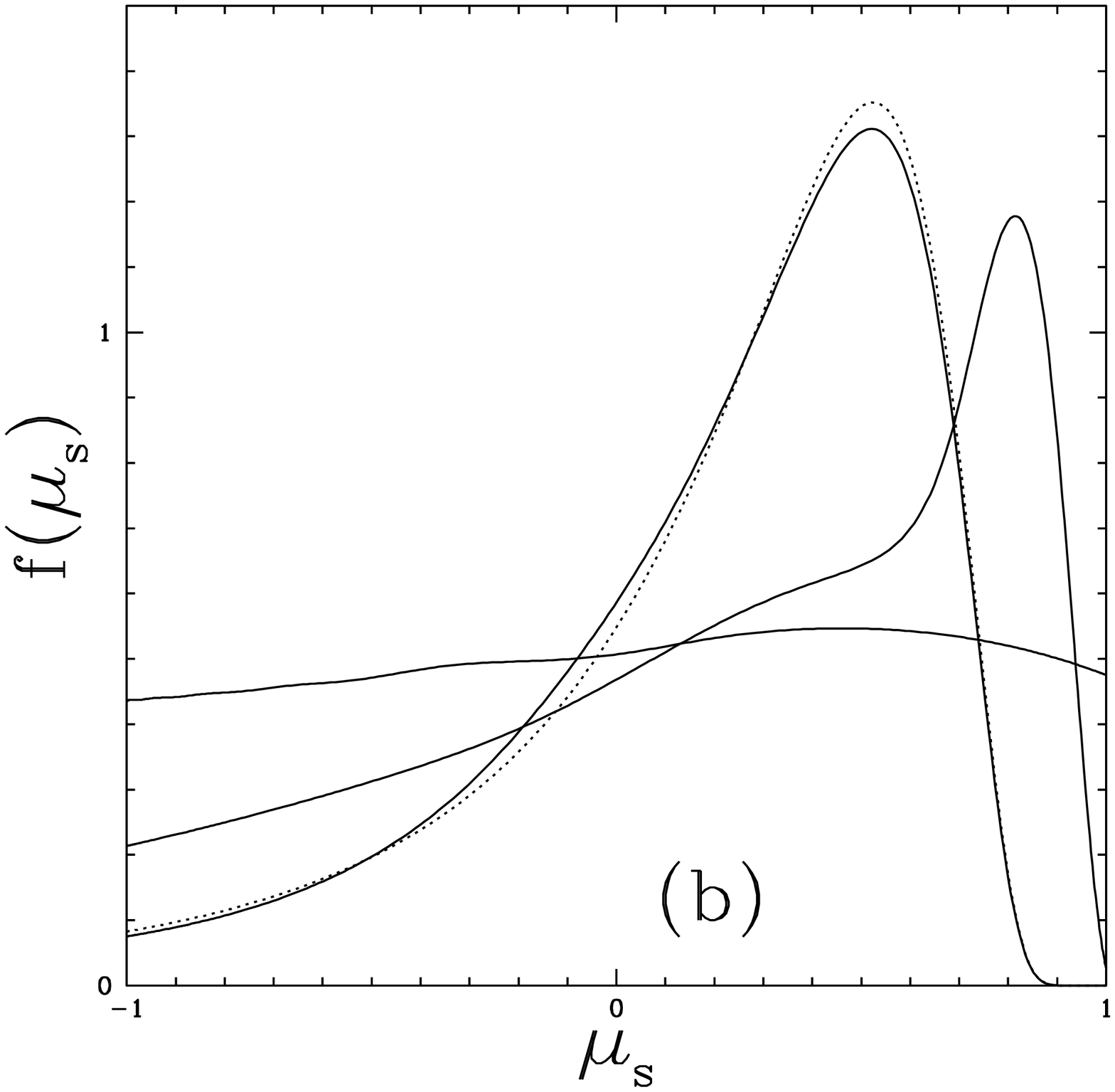}
\caption{\protect\label{anisodist}
The distribution function at the shock front 
for anisotropic diffusion 
according to Eq.~(\protect\ref{tangled}) with $\lpar/\lperp=0.1$ 
(a) as seen in the downstream rest frame
as a function of the direction 
angle $\mu_+$ and
(b) as seen in the shock rest frame, as a function of 
the direction angle $\mu_{\rm s}$. Curves are plotted for 
three different values of the upstream speed: $\Gamma_-u_-=0.1$, $1$, 
and $10$. The maximum value of $f$ rises monotonically with $u_-$.
For $\Gamma_-u_-\ge10$, there is no further discernable change in the 
distribution. In (b) the dotted line shows, for comparison, the approximate 
expression given in Eq.~(\ref{asympdist}), using $\Gamma_-u_-=10$, and the 
appropriate value of $s$. The jump conditions used are those of a strong shock with the J\"uttner/Synge equation of state downstream}
\end{figure}

\begin{figure}
\plotone{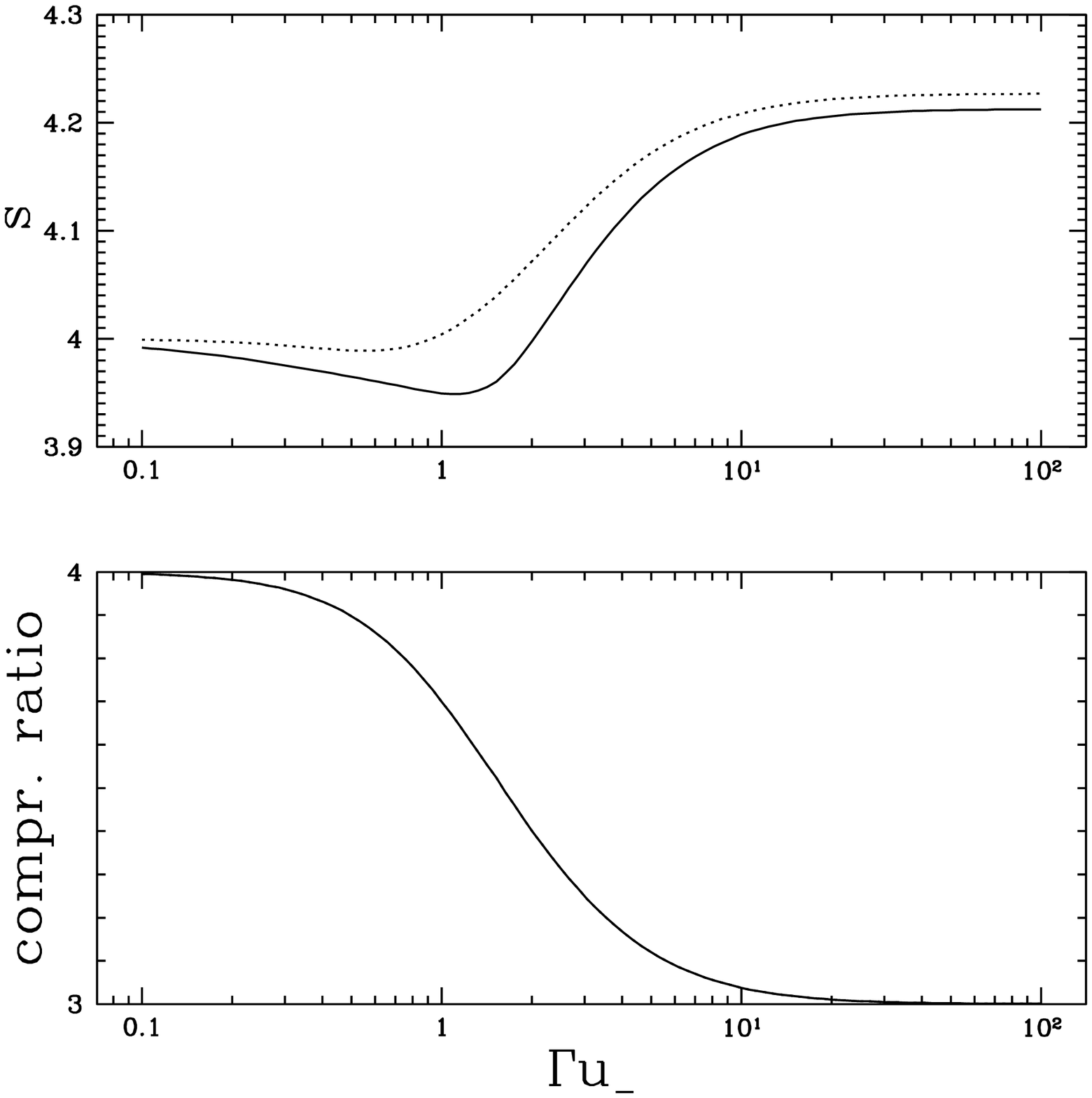}
\caption{\protect\label{anisotropfig}
The power-law index for anisotropic diffusion at 
a strong hydrodynamic shock front. The dotted line shows the 
result for isotropic diffusion, the solid line that for 
diffusion in a tangled field with short correlation length along the direction
of the shock normal, according to Eq.~(\protect\ref{tangled})
with $\lpar/\lperp=0.1$. The lower panel shows
the compression ratio of the shock. 
The jumps conditions are the same as those used for Fig.~5
}
\end{figure}

\subsection{Anisotropic diffusion in direction angle}
\label{anisotropicdiff}
The assumption implicit in the above computations is that the magnetic field
is tangled on short length scales, such that the particle motion can be
described as diffusion in direction angle (the angle between the shock normal
and the particle velocity). The diffusion coefficient 
$D_{\mu\mu}=(1-\mu^2)D$, with $D$ constant, applies in a magnetic field 
which has no preferred direction in
space.  

On the other hand, it is reasonable to suppose that the turbulence at a relativistic shock does 
have a preferred direction -- that of the shock normal -- leading us to expect 
anisotropic diffusion as discussed in Sect.~\ref{transport}.
Figure~\ref{anisodist} shows the angular distribution at the shock front when
diffusion proceeds according to Eq.~(\ref{tangled}) with
$\lpar/\lperp=0.1$. In this figure we also plot the approximate expression for the 
distribution given in Eq.~(\ref{asympdist}), evaluated using $s$ from the full 
numerical solution. Although the distribution at $\Gamma_-u_-$ has converged in the sense 
that it does not change appreciably for higher speeds, it remains significantly different from the 
approximate expression. This emphasizes the necessity of including higher eigenfunctions in the
treatment of highly anisotropic downstream diffusion coefficients.  
Using the diffusion coefficient in
Eq.~(\ref{tangled}) enables particles at small $\mu$
to diffuse rapidly in angle. This leads to a
marked change in the form of the distribution which is especially pronounced
at intermediate shock speeds. Instead of diffusing preferentially along the
shock front, particles at these shocks are rapidly moved to directions closer
to the shock normal. 

This results in an enhancement of the average momentum gain per crossing and
recrossing cycle, but also increases the probability that a particle escapes
when on the downstream side of the shock. 
Results for the power-law index using anisotropic diffusion at a strong
shock are presented in Fig.~\ref{anisotropfig} and show harder values than for 
the isotropic
scattering case. The largest effect again appears 
for mildly relativistic shock speeds, but the difference between
isotropic and anisotropic scattering is not dramatic
amounting to a decrease in $s$ of at most $0.06$. In
the ultrarelativistic limit, there is a small but noticeable effect,
corresponding to a decrease of about $0.02$.

\section{Discussion}
\label{discuss}

We present semi-analytic computations of the power-law index of particles
undergoing first order Fermi acceleration at ultrarelativistic shocks.
In the limit of high shock speed, the index tends to a value of 
$s=4.23$ when the magnetic field is dynamically unimportant. This agrees with 
the results of Monte-Carlo simulations of both highly relativistic shocks
\citep{bednarzostrowski98,achterbergetal00} and of the limiting case
of ultrarelativistic shocks \citep{gallantetal00}.  The result is 
independent of the equation of state of the plasma, since, 
in the ultrarelativistic limit, the compression ratio 
of all hydrodynamic shocks tends to $3$. If the effects of cooling are
negligible, this distribution gives rise to
synchrotron radiation with a spectral index $\alpha=0.62$ (where 
$\alpha=-\diff \ln F_\nu/\diff\nu$, $F_\nu$ is the 
flux of radiation and $\nu$ the frequency). On the other hand,
if the radiative cooling time is short compared to the escape time from the
source the observed spectrum steepens to $\alpha=1.12$.
Distributions with spectral indices compatible with this value have
been identified as responsible for the synchrotron emission in various objects
thought to contain highly relativistic shock fronts: gamma-ray
bursts afterglows \citep{galamaetal98}, the Crab Nebula 
\citep{kennelcoroniti84b} and the Blazar Mkn~501
\citep{krawczynskietal00}.

Physically, the spectrum is determined by the extent to which particles
crossing and re-crossing the shock front form a population whose velocity
vectors are beamed predominantly along the shock surface. This
\lq capturing\rq\ by the shock has two effects: on
the one hand it reduces the probability that the particles are swept away from
the shock front, reducing their escape probability per cycle, but, on the other
hand, it also reduces the mean energy gained per cycle 
\citep{gallantachterberg99}. For an isotropically distributed incoming
population, the relative energy gain upon crossing and re-crossing is
approximately $\Gamma_-^2$ \citep{vietri95}, but for particles which
undergo many crossings, this decreases to become of the order of unity. 

As in the case of mildly relativistic shocks, anisotropic transport  
may in principle affect the result. We compute an example
in which scattering is due to a tangled magnetic field which is not isotropic,
but has a shorter correlation length along the shock normal.
The result is a relatively minor correction to the 
power-law index.  
The \lq universality\rq\ of the index $4.23$ can, however, be broken if 
the shock front moves into a strongly magnetized plasma. We present results
which indicate that for a value of $\sigma=10^{-2}$ the asymptotic index for 
ultrarelativistic shocks steepens to $s=4.30$.
This is mainly a result of the decreased
compression ratio of such shocks, a phenomenon which 
has been described for mildly
relativistic shocks by \citet{ballardheavens91} and
may be of importance for acceleration at shocks in pulsar driven winds.

\acknowledgments
This work was supported by the European 
Commission under the TMR Programme, contract number FMRX-CT98-0168.

\appendix
\section{Eigenfunctions}
Using the Pr\"ufer transformation:
\eqb
Q(\mu)&=&\exp[w(\mu)]\sin\Theta(\mu)\\
D(\mu)(1-\mu^2){\diff Q\over\diff\mu}&=&\exp[w(\mu)]\cos\Theta(\mu)\\
\eqe
the eigenvalue problem Eq.~(\ref{eigvalprob}) 
can be transformed into two first order differential equations:
\eqb
\Theta'&=&{\cos^2\Theta\over D(\mu)(1-\mu^2)}-\Lambda(u+\mu)\sin^2\Theta
\label{pruefera}
\\
w'&=&\sin\Theta\cos\Theta\left[{1\over D(\mu)(1-\mu^2)}+\Lambda(u+\mu)\right]
\label{prueferb}
\eqe
together with the boundary conditions that $\Theta$ and $w$ are regular at
$\mu=\pm1$, i.e.
\[
\left.
\begin{array}{r@{\,=\,}l}
\cos\Theta&0\\
\Theta'&\Lambda(1-u)\\
w'&-{\Lambda(1-u)/2 D(-1)}
\end{array}
\right\rbrace {\rm at }\ \mu=-1
\]
and
\[
\left.
\begin{array}{r@{\,=\,}l}
\cos\Theta&0\\
\Theta'&-\Lambda(1+u)\\
w'&-{\Lambda(1+u)/2 D(1)}
\end{array}
\right\rbrace {\rm at }\ \mu=+1
\]
There are two advantages to this method. Firstly, it is necessary to solve 
only a single first-order differential equation to find the eigenvalues. 
Secondly, taking $\Theta=\pi/2$ at $\mu=-1$, 
each individual eigenvalue $\Lambda_n$ is found by solving a unique
boundary value problem with $\Theta=(2n-1)\pi/2$ at 
$\mu=+1$. This is easily achieved using a shooting method to match the 
solutions at the point $\mu=-u$.

For ultrarelativistic flows, one may define the small parameter
$\epsilon=1-u$ and the ODE's (\ref{pruefera}) and (\ref{prueferb}) 
possess two boundary layers.
For $\lambda=\epsilon^2\Lambda\sim O(1)$, these have 
thickness $O(\epsilon)$  at $\mu=-1$ and $O(\epsilon^2)$ at $\mu=+1$. In terms 
of the stretched variable describing the left-hand layer $y=(1+\mu)/\epsilon$,
the zeroth order equation in $\epsilon$ reads:
\eqb
{\diff\Theta\over\diff y}&=&{\cos^2\Theta\over 2y D(-1)}- 
\lambda\sin^2\Theta(y-1)
\label{zeroth}
\eqe
with boundary conditions $\cos\Theta=0$ at $y=0$ and $\sin\Theta=0$ 
at $y\rightarrow\infty$. This equation can be transformed into the 
linear second order differential equation given in
Eq.~(A4) of \citet{kirkschneider89}, which has the
solutions given in Eqs.~(\ref{asymptoticeigf}) 
and (\ref{asymptoticeigv}). 
However, especially for high-order eigenfunctions, it is 
more convenient to evaluate the solution 
by numerical integration of Eqs.~(\ref{pruefera})
and (\ref{prueferb}), taking
the approximate boundary condition $\Theta=n\pi$, at $\mu=1-O(\epsilon)$.

\end{document}